\documentclass[conference,10pt]{IEEEtran}
\usepackage[]{graphicx}
\usepackage[]{amsfonts,amsmath, amssymb,calrsfs}
\usepackage[mathscr]{eucal}
\usepackage{cuted}
\usepackage{lipsum,amsmath,multicol}

\def\eq#1{Eq.~\eqref{eq:#1}}
\def\fig#1{Fig.~\ref{fig:#1}}
\DeclareMathAlphabet{\mathpzc}{OT1}{pzc}{m}{it}

\def\ket#1{| #1 \rangle}

\def\cC{\mathcal{C}}
\def\cE{\mathcal{E}}
\def\cF{\mathcal{F}}

\def\cL{\mathcal{L}}

\def\cO{\mathcal{O}}

\def\cS{\mathcal{S}}

\def\cZ{\mathcal{Z}}

\DeclareMathOperator*{\argmax}{\rm{argmax}}

\DeclareMathAlphabet{\mathpzc}{OT1}{pzc}{m}{it}

\begin{document}

\title{A renormalization group decoding algorithm for topological quantum codes}

\author{
\authorblockN{Guillaume Duclos-Cianci and David Poulin}
\authorblockA{D\'epartement de Physique, Universit\'e de Sherbrooke, Qu\'ebec, Canada \\
Emails: Guillaume.Duclos-Cianci@USherbrooke.ca, David.Poulin@USherbrooke.ca}
}

\maketitle

\begin{abstract}
Topological quantum error-correcting codes are defined by geometrically local checks on a two-dimensional lattice of quantum bits (qubits), making them particularly well suited for fault-tolerant quantum information processing. Here, we present a decoding algorithm for topological codes that is faster than previously known algorithms and applies to a wider class of topological codes. Our algorithm makes use of two methods inspired from statistical physics: renormalization groups and mean-field approximations. First, the topological code is approximated by a concatenated block code that can be efficiently decoded. To improve this approximation, additional consistency conditions are imposed between the blocks, and are solved by a belief propagation algorithm. 
\end{abstract}

\IEEEpeerreviewmaketitle

\section{Introduction}

Topological quantum error-correcting codes (TQECC) \cite{DKLP02a,Kit03a,BM07a} are defined on two-dimensional lattices of qubits with geometrically local parity checks. Thus, they are a form of quantum LDPC codes with an additional locality requirement imposed to their Tanner graph. To appreciate the importance of this feature, recall that in quantum mechanics, measuring a qubit alters its state. To detect errors, it is not possible to measure each qubit separately and verify that they satisfy all the check conditions---like it is done classically---without destroying the encoded information. Instead, it is necessary to perform a collective measurement on all the qubits involved in a given check, which requires having the qubits physically interact with each other, or with a mediator system. Thus, having local checks is an extremely important feature that explains---together with the possibility of implementing some gates topologically \cite{Kit03a,FKLW02a,RH07a,RHG07a,BR07a}---the growing interest in topological quantum codes.

The prominent example of TQECC is Kitaev's toric code family \cite{DKLP02a,Kit03a} that we define below. For these codes, defined on a toric qubit lattice, errors in the same homology class have the same effect on the encoded information. Thus,  maximum-likelihood decoding consists in identifying the lowest weight homology class of equivalent errors.  Previously, a decoding algorithm based on perfect matching \cite{E65a} was proposed which identifies the lowest weight error, ignoring the equivalence relation set by homology \cite{DKLP02a}. The complexity of this algorithm is quite prohibitive, $\cO(\ell^6)$ where $\ell$ is the linear size of the lattice. Other topological codes \cite{BM07a} had no known efficient decoding algorithm. 

In \cite{DP10a} we presented a new decoding algorithm for TQECC. The essential idea of this algorithm borrows from the renormalization group method of statistical physics. Intuitively, we can think of a TQECC on a lattice of linear size $\ell = 2^c$ as consisting of $c$ levels of concatenation of a TQECC on a lattice of linear size 2. Concatenated quantum codes can be decoded efficiently by a recursive algorithm \cite{Pou06b}. Starting from an error model characterizing the channel, each $2\times 2$ lattice is soft-decoded, producing an effective ``renormalized" error model on its logical qubits. This error model is passed to the next level of concatenation, and we recurse. The recursion ends after $c = \log \ell$ iterations, where it outputs a probability vector describing the encoded information. Each round involves decoding at most $\ell^2$ constant size TQECC, so the overall complexity is $\cO(\ell^2\log\ell)$ and can easily be parallelized for a total runtime of $\cO(\log\ell)$. 

Because TQECC are not truly concatenated codes, the intuition explained above cannot be turned into a rigorous method, and some approximations are necessary. In this paper, we give a detailed presentation of the approximation techniques used in Ref. \cite{DP10a} and present some results obtained from our method. 

\section{Kitaev's toric code}

The state of a collection of $n$ qubits can be specified by a vector $\ket\psi$ in the complex Hilbert space $(\mathbb{C}^2)^{\otimes n} = \mathbb{C}^2\otimes\mathbb{C}^2\otimes\ldots\otimes\mathbb{C}^2$. Each vector space $\mathbb{C}^2$ in this tensor decomposition is associated to a qubit. A code on $n$ qubits is a subspace of $(\mathbb{C}^2)^{\otimes n}$. For Kitaev's code---like all stabilizer codes---this subspace is specified by a set of mutually commuting operators 
that play a role similar to the rows of a parity-check matrix. 

To define these operators, it is convenient to display the qubits on a regular square lattice with periodic boundary conditions, i.e. with the topology of a torus. There is one qubit on each {\em edge} of the lattice, for a total of $n=2\ell^2$ qubits for a $\ell\times\ell$ lattice. For each site $s$ (vertex) of the lattice, we define a site operator $A_s = \bigotimes_{e\in s} X_e$ and for each plaquette $p$ (site of the dual lattice), we define a plaquette operator $B_p = \bigotimes_{e\in p} Z_e$. These definitions use the Pauli matrices 
$$
X = \left(\begin{array}{cc} 0 & 1\\ 1&0\end{array}\right)\!,\ 
Y = \left(\begin{array}{cc} 0 & -i\\ i&0\end{array}\right)\!,\ {\rm and\ }
Z = \left(\begin{array}{cc} 1 & 0\\ 0&-1\end{array}\right),
$$
and we use $X_e$ to denote the Pauli operator $X$ acting on the qubit located on edge $e$, i.e. $X_e = I\otimes I\otimes\ldots\otimes X\otimes\ldots\otimes I$ where $X$ appears at position $e$ and $I$ denotes the $2\times 2$ identity matrix.  The notation $e\in s$ denotes the set of edges $e$ adjacent to site $s$, and similarly for $v\in p$.

Because $ZX = iY$,  the Pauli operators $X_e$ and $Z_e$, $e=1,\ldots,n$ form a group under multiplication, the Pauli group of $n$ qubits $\langle i, X_j, Z_j \rangle$. Every element in this group squares to the identity (modulo a phase, that we henceforth omit). These generators obey canonical commutation relations: all pairs of generators commute, except $X_e$ and $Z_e$ that anti-commute. It follows from these relations that the $A_s$ and $B_p$ all mutually commute. The commutation of the $A_s$ among themselves is trivial since they are all made up of $X$ matrices, and similarly for the commutation of the $B_p$ among themselves. The commutation of a $A_s$ with a $B_p$ follows from the fact that a site and a plaquette either have no common edge, or they have two. Both cases imply an {\em even} number of anti-commuting operators, so they commute.  

The code is now defined as the subspace
\begin{equation}
\cC = \{\ket\psi \in ({\mathbb{C}}^2)^{\otimes n} : A_s \ket\psi = B_p\ket\psi = \ket\psi \quad\forall s,p\}.
\label{eq:code}
\end{equation}
This is an eigenvalue equation, and because the $A_s$ and $B_p$ mutually commute, they indeed share common eigenvectors. In this definition of the code $\cC$, the operators $A_s$ and $B_p$ play a role analogous to the rows of a parity check matrix in a classical linear code in the sense that they impose (local) constraints on the codewords. 

We will consider errors $F$ from the Pauli group. In particular, we will be interested in the {\em depolarizing channel}---the natural generalization of the binary symmetric channel to the quantum setting---for which each qubit is left unchanged with probability $1-p$, or is affected by $X$, $Y$ or $Z$ each with probability $\frac p3$. In other words, all errors $F$ from the $n$-qubit Pauli group (modulo phase) are permitted, and the probability of a given error is 
\begin{equation}
P(F) = (1-p)^{n-|F|}\frac p3^{|F|},
\label{eq:depol}
\end{equation}
where the weight of $F$, denoted $|F|$, is the number of tensor factors on which it differs from the identity (generalizing the Hamming weight).

When an error $F$ occurs on a code state $\ket\psi$, the resulting state $F\ket\psi$ will in general no longer be a $+1$ eigenstate of the star and plaquette operators: $F\ket\psi$ is a $+1$ eigenstate of $A_s$ when $F$ and $A_s$ commute, and it is a $-1$ eigenstate when they anti-commute, and similarly for $B_p$. By measuring each star and plaquette operators, we obtain a list of $c_s$ and $c_p = \pm 1$ forming the error syndrome $c = (c_p,c_s)$. Given these, error correction can proceed by identifying the most likely error compatible with the syndrome, $F_{ML} = \argmax_{F \in \cF_c} P(F)$ where $\cF_c = \{F: FA_s = c_s A_sF{\rm \ and\ } FB_p = c_p B_pF\}$, and applying this error again to correct it (since $F^2 = \pm I$). As we will now explain, finding the most likely error is not the optimal decoding strategy in quantum mechanics.  

\begin{figure}
\begin{center}
\includegraphics[scale=0.7]{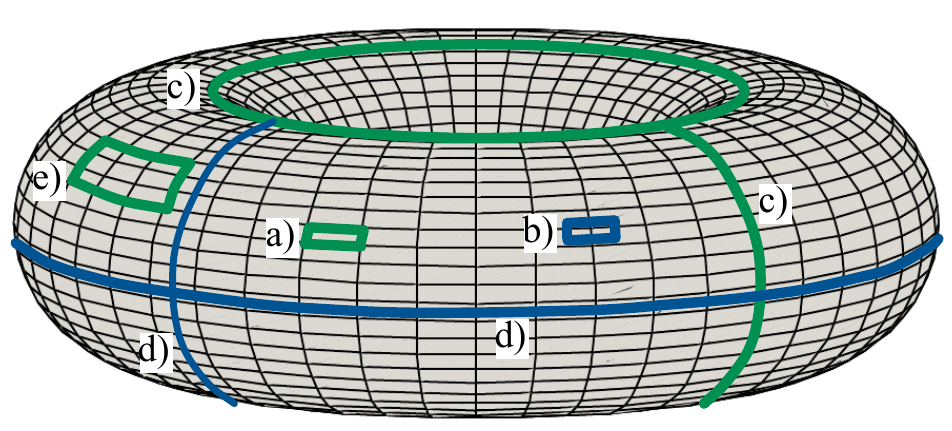}
\caption{A green line indicates the presence of a $Z$ operator on qubit associated to the edge, and blue line indicates a $X$. All $Z$-type operators are strings on the lattice while $X$-type operators are strings on the dual lattice. Trivial loops on the a) lattice and b) dual lattice corresponding to $B_p$ and $A_s$ respectively. Loops with non-trivial homology c) on the lattice and d) on the dual lattice corresponding to the 4 generators of $\cL$. e) A trivial loop obtained by product of elementary trivial loops.}
\label{fig:homology}
\end{center}
\end{figure}

The property that $A_s$ and $B_p$ leave each code state invariant, c.f. \eq{code}, is inherited by the Abelian group generated by them, called the {\em stabilizer group} $\cS = \langle A_s,B_p\rangle$. This induces an equivalence relation between operators on $(\mathbb{C}^2)^{\otimes n}$. If two operators $F$ and $F'$ differ by an element of $\cS$, for instance if $F' = FB_4A_7$, then they will have an identical effect on $\cC$ by definition. Notice that the plaquette operators $B_p$ are elementary loops of $Z$ on the lattice, so their products generate the group of homologically trivial loops on the torus, see \fig{homology}. Likewise, the star operators generate the group of homologically trivial loops of $X$ on the dual lattice. We conclude that two operators $F$ and $F'$ have the same effect on the code if $FF'$ contains only homologically trivial loops, in which case we say that $F$ and $F'$ are homologically equivalent. 

On the other hand, there are 4 independent operators that map $\cC$ to itself (i.e. commute with all star and plaquette operators) but do not belong to $\cS$; they correspond to the homologically non-trivial loops of $Z$ and $X$ around the hole and the body of the torus, see \fig{homology}. They generate a group with $2^4=16$ operators called the {\em logical Pauli operators} $\cL$, and are associated with two encoded qubits. Because $L$ and $LS$ are equivalent for $S\in \cS$, the choice of the 4 generators of $\cL$ depicted on \fig{homology} is to some extent arbitrary, only the homology class of the operators matters. Thus optimal decoding consists in identifying the most likely class of homologically equivalent errors, in other words the most likely logical operator 
\begin{equation}
L_{ML} = \argmax_{L\in \cL} \sum_{S\in\cS} P(F = LST(c_s,c_p))
\label{eq:MLH}
\end{equation}
where $T(c_s,c_p)$ is any reference error compatible with the error syndrome $c_s,c_p$. Error correction is completed by applying $L_{ML}T(c_s,c_p)$. Although \eq{MLH} is expressed in terms of a specific choice of generators of $\cL$ and a specific reference error $T(c_s,c_p)$ that are not topological invariants, the sum over $\cS$ makes the homology class of the correction operator $L_{ML}T(c_s,c_p)$ independent of these choices. 

\section{Renormalization algorithm}

\begin{figure}
\begin{center}
\includegraphics[scale=0.65]{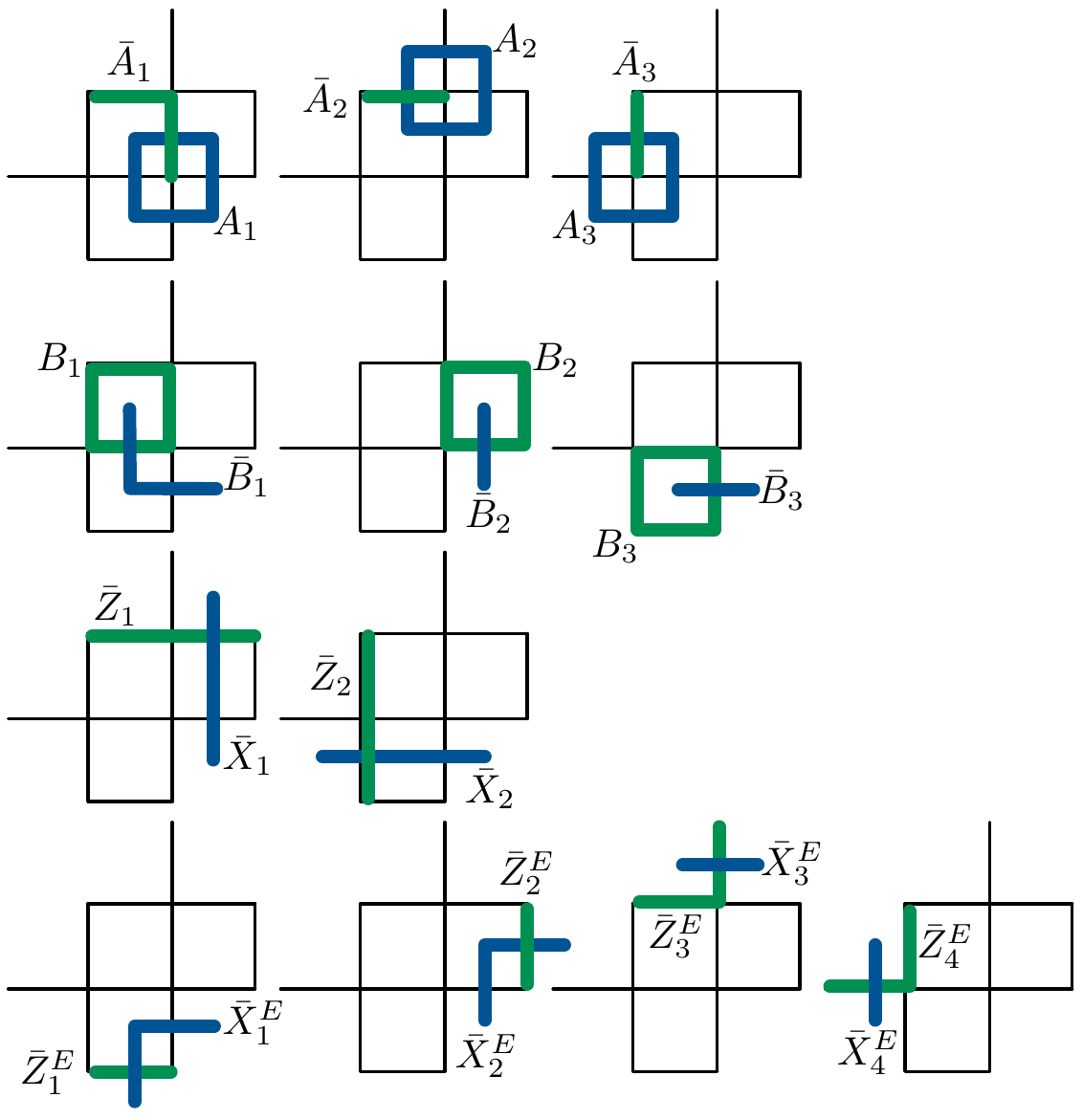}
\caption{Definition of the Pauli operator basis used in the renormalization algorithm, color scheme as in \fig{homology}. This operator basis obeys canonical commutation relations: any two operators commute, except if they are illustrated on the same unit cell in which case they anti-commute. Operators defined in the first two rows are stabilizer generators and pure errors, on the third row are logical operators, and on the last row are edge operators.}
\label{fig:basis}
\end{center}
\end{figure}

\begin{figure}
\begin{center}
\includegraphics[scale=0.65]{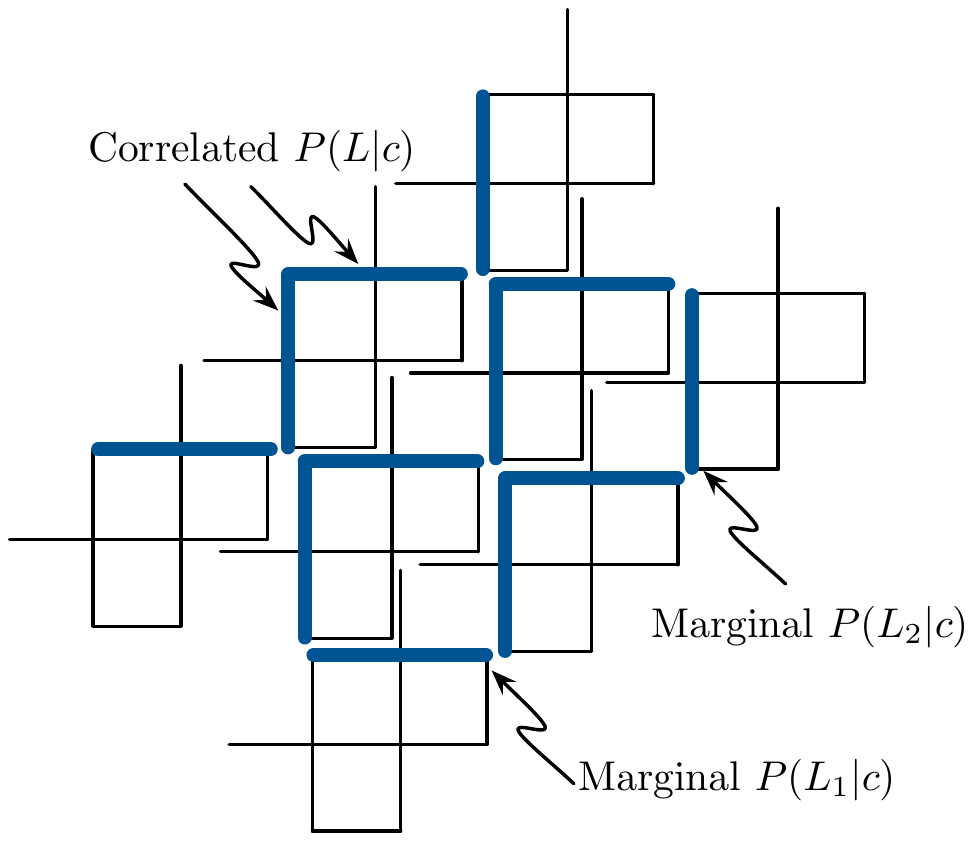}
\caption{The lattice is broken into overlapping unit cells (black) that are shifted on the figure for illustration purposes. The logical qubits from 8 bare unit cells are joined to construct a larger renormalized unit cell (blue). The error probability $P(F)$ for this larger unit cell is obtained by combining the underlying  logical operator distributions conditioned on the error syndromes of each cell. When both logical qubits of a unit cell participate in the construction of the larger cell, the induced error model on these qubits is correlated as illustrated on the figure. When only one logical operator participates in the construction, we consider the marginal distribution on that logical  qubit.}
\label{fig:glue}
\end{center}
\end{figure}

The optimal recovery scheme described above is in general very hard to achieve computationally. In particular, summing over the entire group of loops $\cS$ is a formidable task. To circumvent this difficulty, we will attempt to ``divide and conquer"  using an approach inspired by the renormalization group method of statistical mechanics. We break the lattice into overlapping ``unit cells" as illustrated in \fig{glue}. Each of these cells contains 12 edges, and hence 12 qubits. The choice of this unit cell is somewhat arbitrary, but we will stick to this particular example for concreteness. This cell encloses 6 stabilizer generators in total, three plaquettes and three sites (shown on the first two rows of \fig{basis}). We can use these stabilizers to define a (small) error-correcting code---a surface code \cite{BK98a,FM01a}, open boundary version of Kitaev's toric code. 

To describe the algorithm, it is convenient to choose a set of generators for the Pauli group on the lattice. Our choice is defined in \fig{basis}. Note that these operators obey canonical commutation relations, the same as the one of $X_e$ and $Z_e$: any two of these operators commute except for the two illustrated on the same unit cell that anti-commute. We also group these operators into three categories that play different roles in our algorithm, as we now explain.

The stabilizer generators $A_s$ and $B_p$ play the same double role as explained in the previous section: they are measured to read-out the error syndrome, and they induce an equivalence relation between operators. They generate the stabilizer group $\cS$ containing $2^6 = 64$ elements. Their conjugate partners $\bar A_s$ and $\bar B_p$, that we call ``pure errors", are used to construct the reference error $T(c_s,c_p)$ appearing in \eq{MLH} in a systematic way. Indeed, it follows from the canonical commutation relations that the operator 
\begin{equation}
T(c_s,c_p) = \prod_{s:c_s=-1} \bar A_s\prod_{p: c_p=-1} \bar B_p
\label{eq:ref_error}
\end{equation}
is an error with syndrome $c_s,c_p$. 

The four logical operators are representative of the homologically non-trivial loops (although they don't look like loops, they are strings with no ends on the lattice, which is the definition of a loop). They generate the logical group $\cL$ containing $2^4 = 16$ elements. In our recursive decoding algorithm, these four generators will be mapped onto single qubit operators on a renormalized lattice of half the linear size. Hence, our goal is to assign probabilities to these logical operators, that will serve as an effective channel for the following recursion. 

Finally, the four pairs of edge operators are needed to complete the set of generators, and correspond roughly (but see below) to qubits that are shared between neighboring unit cells. They generate the edge group $\cE$ containing $4^4 = 256$ elements. Thus, they will be used to ``glue" neighboring unit cells into a renormalized unit cell used in the following recursion of the algorithm.

With these definitions in place, we can describe the elementary step of our renormalization procedure on a given cell, which consists in computing a conditional probability distribution. We are given a list of syndromes $(c_s,c_p) \in \{-1,1\}$ associated to the six stabilizer generators on the unit cell. We collectively denote these syndromes by $c$. We are also given a  probability $P(F)$ of errors $F$ contained on the cell. In the first step of the recursion, this probability is set by the channel, e.g. \eq{depol}, and in the $k$'th step it is given by the output of step $k-1$. From these, we compute the joint probability on the logical and edge group elements conditioned on the syndrome
\begin{equation}
P(L,E|c) = \frac 1\cZ \sum_{S\in \cS} P(F = LEST(c)),
\label{eq:conditional}
\end{equation}
where $T(c)$ is the reference error defined at \eq{ref_error}, and $\cZ$ is a normalization factor. Note that this equation is very similar to the definition of the maximum-likelihood decoder \eq{MLH}, except that we must now include edge operators and we do not commit to a hard decision but instead keep the entire probability vector. This procedure is repeated for all the unit cells of the lattice (this can be done in parallel, so in a constant amount of time). 

We will be using different marginals of this probability. For each cell, we can define the marginal probability on the logical and edge operators by $P(L|c) = \sum_{E\in \cE} P(L,E|c)$ and $P(E|c) = \sum_{L\in \cL} P(L,E|c)$ respectively. Notice that, up to multiplication by $\bar A_s$ and $\bar B_p$, the edge operators $E$ are supported only on those qubits that are shared between neighboring unit cells. For instance, the product $\bar X_1^E\bar B_3$ is the operator $X$ acting on the bottom shared qubit, see \fig{basis}. Moreover, because the pure error component of the error is determined by $c$, c.f.  \eq{ref_error}, we can directly interpret the conditional probability on the edge operators as a probability distribution for Pauli operators on the shared qubits. This feature will be important when we describe how we glue unit cells together using belief propagation. 

Each of these marginal probabilities can be broken down even further. Remember that the logical operators $\cL$ are generated by two pairs of canonically conjugated operators $\bar X_{j}$ and $\bar Z_{j}$, $j=1,2$, representing the two logical qubits, c.f. \fig{basis}. Thus, every operator in $\cL$ can be written as $L = L_1  L_2$ with $L_j \in \cL_j = \langle \bar X_j, \bar Z_j\rangle$. We can consider the marginal on one of the two logical qubits by summing over the value of the other variable, e.g. $P(L_1|c) = \sum_{L_2\in \cL_2} P(L=L_1L_2|c)$. Likewise, the edge operators are generated by four canonical pairs of edge operators $\overline X_j^E$ and $\overline Z_j^E$, each encoding an edge qubit, c.f. \fig{basis}. We can express any edge operator as $E = E_1E_2E_3E_4$ with $E_j \in \cE_j = \langle \bar X_j^E, \bar Z_j^E \rangle$, and consider marginals such as $P(E_1|c) = \sum_{E' \in \cE_2\times\cE_3\times\cE_4} P(E=E_1E'|c)$. Following the discussion above, these marginal probabilities can be directly interpreted as probability of the Pauli operators acting on the shared qubits.  For instance, $P(E_1|c)$ is the error probability of the bottom shared qubit conditioned on the error syndrome of the unit cell.

To complete one step of the renormalization procedure, we join the logical operators from 8 ``bare" unit cells into a larger renormalized unit cell as shown on \fig{glue}. In a first approximation, we can use the probabilities $P(L|c)$ from each bare unit cell to assign an effective error model to the renormalized unit cell. Note that the conditional probability $P(L|c)$ defined above is a joint probability distribution on two logical qubits. As a consequence, the renormalized error model can have correlated errors between neighboring qubits. This is not a problem when the two qubits appear in the same renormalized unit cell. We can take these correlations into account in our definition of the renormalized error model $P(F)$. However, when these two correlated qubits belong to two distinct renormalized unit cells, it is not possible to keep track of these correlations efficiently. In those cases, we use the appropriate marginal distributions on each qubit to define the renormalized error model, see \fig{glue}. In other words, we replace $P(L_1,L_2|c)$ by $P(L_1|c)P(L_2|c)$. Ignoring some of the correlations between the renormalized qubits is an approximation used to make our scheme efficient.

\begin{figure}
\begin{center}
\includegraphics[scale=0.65]{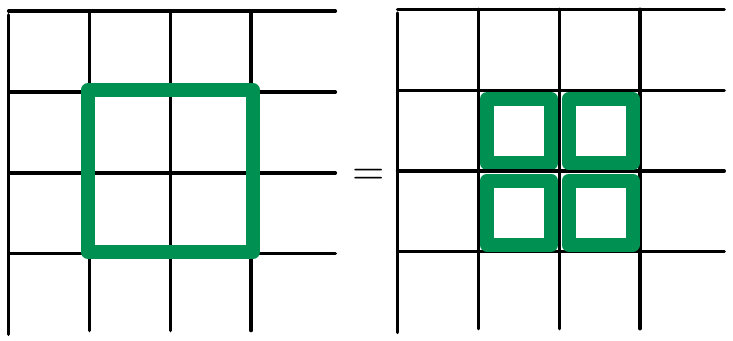}
\caption{The syndrome $c_p$ associated to a renormalized plaquette operator is equal to the binary product of the four smaller plaquette operators contained in it. This follows from the fact that all these operators commute, and the product of the four operators is equal to the renormalized operator. The same holds for the star operators.}
\label{fig:stabilizer}
\end{center}
\end{figure}

The same procedure can now be executed on each renormalized unit cell. We can compute the various probabilities conditioned on the error syndromes contained in each cell. Note that the renormalized star and plaquette operators each act on 8 bare qubits, spoiling the locality feature of Kitaev's toric code. However, this renormalization of the star and plaquette operators is only for the purpose of presenting the decoding algorithm. The error syndrome associated to the renormalized stabilizer generators \cite{C04a} can be obtained by measuring the original 4-qubit star and plaquette operators and multiplying their outcomes as shown in \fig{stabilizer}. 

At the last iteration of this renormalization procedure, we obtain a probability distribution over the logical operators of the encoded qubits, completing the soft decoding procedure. The operator with the largest probability can be selected to implement the correction.

\section{Belief propagation}

\begin{figure}
\begin{center}
\includegraphics[scale=0.65]{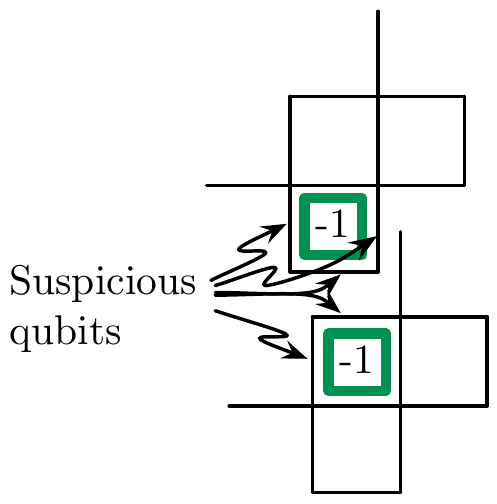}
\caption{Two adjacent unit cells each have one $c_p = -1$ as illustrated, and all the other error syndromes are $+1$. For the top cell, three errors are equally likely to have caused this syndrome: an $X$ error on the qubit to the left, the bottom, or the right of the plaquette operator. Similarly for the bottom cell, the two dominating errors that could have caused the syndrome are an $X$ error to the left and above the plaquette. But in fact, the bottom qubit of the top cell and the top qubit of the bottom cell are actually the same qubit; a shared qubit. This qubit will be assigned a probability roughly $\frac 13$ of having an $X$ error by the top cell, while the bottom cell will assign the probability $\frac 12$, which is inconsistent. On the other hand, because it is doubly suspicious, the probability of this qubit having suffered an $X$ error should be dominating, but decoding each cell independently fails to recognize this.}
\label{fig:defect}
\end{center}
\end{figure}

The unit cells used by the renormalization decoding algorithm overlap in the sense that some qubits are shared between two unit cells. Without these overlaps, each unit cell would contain only 2 complete stabilizer generators ($A_1$ and $B_1$, see \fig{basis}) instead of 6. As a consequence, the number of variables of the renormalized error model would increase by a constant factor at each renormalization step, leading to an exponential blowup. Thus, these shared qubits appear to be necessary. On the other hand, the presence of shared qubits leads to the main approximation of our decoding scheme: a qubit that is shared between two unit cells is treated independently by each one of them as if it were two independent random variables. This can lead to some inconsistencies as illustrated in \fig{defect}. 

To improve on this approximation, we allow unit cells to exchange messages. The purpose of these messages is to update the error model of the shared qubits---or equivalently of the edge operators---conditioned on the syndromes of the immediate neighboring cells. In short, each cell computes the marginal conditional probability of each of its shared qubits (or edge operators, as explained above), and passes the probability associated to a given edge qubit to the cell sharing it. All unit cells can perform this in parallel. Then, each cell reweighs the prior error model of its edge qubits by the incoming messages. Iterating this procedure leads to a probability that is conditionned on the syndromes of an extended neighborhood. 

The procedure can be formalized as a belief propagation algorithm. Let $P(E_q)$ denote the marginal probability assigned to edge qubit $q$ (obtained from the channel model or the output of the previous renormalization step). At round $t$, each cell $C$ outputs $8$ messages $m^{t,\rm{out}}_{C,q}(E_q)$, one for each of its shared qubit\footnote{Each unit cell has 4 edge qubits, and overlaps with one edge qubit of each of its 4 neighboring cell, for a total of 8 shared qubits.} $q$, that correspond to some probability vector on $E_q \in \{I,X,Y,Z\}$. At the first round, these messages are initialized to the uniform distribution. The cell's outgoing messages at time $t$ become its neighbors incoming messages at time $t+1$: if cells $C$ and $C'$ share qubit $q$, $m^{t+1,{\rm in}}_{(C',q)}(E_q) =m^{t,\rm{out}}_{(C,q)}(E_q)$. The message update rule is given by 
\begin{align}
&m^{t,\rm{out}}_{(C,q)}(E_q) \propto \\
&\frac 1 {P(E_q)} \!\!\!\!\!\! \sum_{\substack{S\in\cS_C \\ L\in \cL_C \\ E_{q'}: q'\in C\backslash q }} \!\!\!\!\!\! P\Big(F=\prod_{q'\in C}E_{q'}LST(c)\Big)\prod_{q'\in C\backslash q} m^{t,\rm{in}}_{(C,q')}(E_{q'}) \nonumber
\end{align}
where notation such as $\cS_C$ denotes the stabilizer group generated by the 6 generators enclosed in $C$, the notation $q\in C$ stands for the set of edge qubits contained in $C$, and finally $C\backslash q$ denotes all the edge qubits in $C$ except $q$. 

The messages roughly converge to steady distributions after a few iterations of this procedure (we typically use 3, since the graph contains 4-cycles). Once convergence is reached, the renormalization algorithm is executed on each cell as explained in the previous section, but using the incoming messages to the cell to reweigh the prior probabilities on the shared qubits. 

\section{Results}

\begin{figure}
\begin{center}
\includegraphics[scale=0.35]{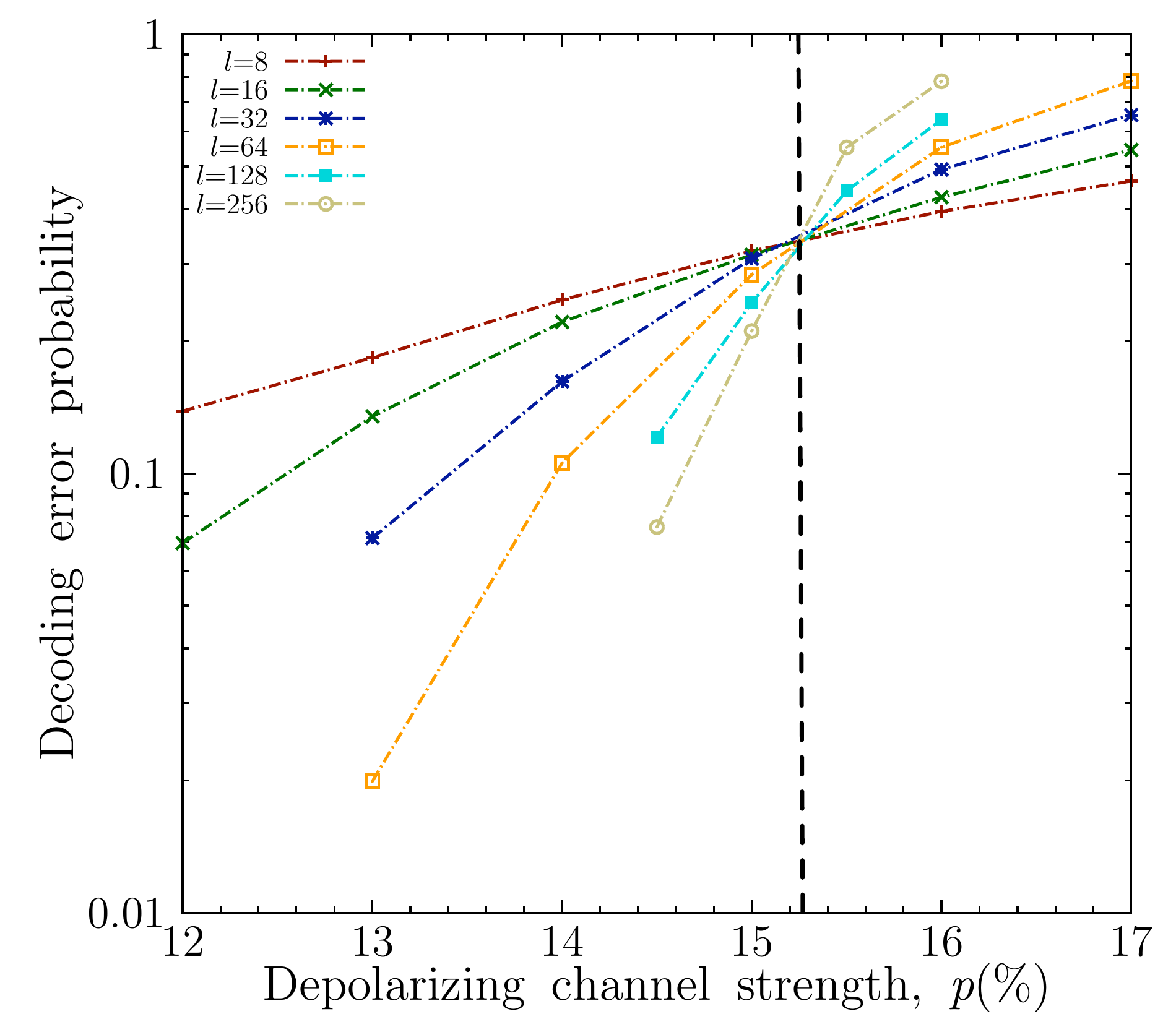}
\caption{Block error probability as a function of the depolarizing strength for toric codes on tori of different linear size $\ell=8,16,\ldots,128$. Our decoding algorithm yields an error threshold of about 15.2\%.}
\label{fig:results}
\end{center}
\end{figure}

We have assessed the performance of our decoder using Monte Carlo sampling. Figure \ref{fig:results} summarizes our results. It shows a clear threshold near the depolarizing probability $p = 15.2\%$, very close to what is achieved by the perfect matching decoder of \cite{DKLP02a}. Thus, we obtain an exponential gain in decoding time without significant performance loss. 

Many modifications can be made to the basic decoding scheme presented here that allow tradeoffs between decoding complexity and error suppression. Some of these extensions were presented in \cite{DP10a}, and in particular they achieved a depolarizing threshold higher than the perfect matching algorithm. We have also adapted our decoder to other noise models, such as the erasure channel, and other topological codes, in particular the color codes of \cite{BM07a}. 

\noindent{\em Acknowledgements---}We thank Jim Harrington, H\'ector Bomb\'in, and Sergey Bravyi for useful conversations. This work was partially funded by NSERC, FQRNT, and MITACS. Computational resources were provided by the R\'eseau qu\'eb\'ecois de calcul de haute performance (RQCHP) and Compute Canada.

\end{document}